\documentclass[conference, compsocconf]{IEEEtran}
\IEEEoverridecommandlockouts
\usepackage{cite}
\usepackage{listings}
\usepackage{amssymb}
\usepackage{bm}
\usepackage{amsmath,amssymb,amsfonts}
\usepackage{algorithm}
\usepackage{algorithmic}
\usepackage{graphicx}
\usepackage{enumerate}
\usepackage{textcomp}
\usepackage{xcolor}
\def\BibTeX{{\rm B\kern-.05em{\sc i\kern-.025em b}\kern-.08em
    T\kern-.1667em\lower.7ex\hbox{E}\kern-.125emX}}

\usepackage{caption}
\begin{document}

\title{Sparse Iterative Solvers Using High-Precision Arithmetic with Quasi Multi-Word Algorithms}

\makeatletter
\newcommand{\linebreakand}{%
  \end{@IEEEauthorhalign}
  \hfill\mbox{}\par
  \mbox{}\hfill\begin{@IEEEauthorhalign}
}
\makeatother

\author{
\IEEEauthorblockN{Daichi Mukunoki}
\IEEEauthorblockA{Information Technology Center\\
Nagoya University\\
Aichi, Japan\\
mukunoki@cc.nagoya-u.ac.jp}
\and
\IEEEauthorblockN{Katsuhisa Ozaki}
\IEEEauthorblockA{Shibaura Institute of Technology\\
Saitama, Japan\\
ozaki@shibaura-it.ac.jp}
}

\maketitle

\begin{abstract}
To obtain accurate results in numerical computation, high-precision arithmetic is a straightforward approach. However, most processors lack hardware support for floating-point formats beyond double precision (FP64). Double-word arithmetic (Dekker 1971) extends precision by using standard floating-point operations to represent numbers with twice the mantissa length. Building on this concept, various multi-word arithmetic methods have been proposed to further increase precision by combining additional words. Simplified variants, known as quasi algorithms, have also been introduced, which trade a certain loss of accuracy for reduced computational cost. In this study, we investigate the performance of quasi algorithms for double- and triple-word arithmetic in sparse iterative solvers based on the Conjugate Gradient method, and compare them with both non-quasi algorithms and standard FP64. We evaluate execution time on an x86 processor, the number of iterations to convergence, and solution accuracy. Although quasi algorithms require appropriate normalization to preserve accuracy -- without it, convergence cannot be achieved -- they can still reduce runtime when normalization is applied correctly, while maintaining accuracy comparable to full multi-word algorithms. In particular, quasi triple-word arithmetic can yield more accurate solutions without significantly increasing execution time relative to double-word arithmetic and its quasi variant. Furthermore, for certain problems, a reduction in iteration count contributes to additional speedup. Thus, quasi triple-word arithmetic can serve as a compelling alternative to conventional double-word arithmetic in sparse iterative solvers.
\end{abstract}


\section{Introduction}
\label{sec:introduction}
To achieve accurate solutions in numerical computation, high-precision floating-point arithmetic is a straightforward approach. However, most processors lack hardware support for precisions beyond double precision (binary64, FP64). Consequently, software-based methods for emulating higher precision have been developed. In 1971, Dekker introduced Double-Word Arithmetic~\cite{dekker1971}, which extends precision by combining two floating-point numbers to effectively double the mantissa length. This method is commonly known as double-double (DD) arithmetic; in this study, we refer to it as DW arithmetic.
Building on this idea, various multi-word arithmetic techniques have been proposed to further enhance precision by using additional words. Simplified variants, known as quasi algorithms, have also been introduced. These algorithms omit the normalization step (described in Section~\ref{sec:algorithm}), resulting in reduced accuracy.
The representative algorithms include:
\begin{itemize}
    \item Double-Word (\textbf{DW}) arithmetic (Dekker 1971~\cite{dekker1971})
    \item Triple-Word (\textbf{TW}) arithmetic (Fabino et al. 2019~\cite{10.1109/TC.2019.2918451})
    \item Quadruple-Word (\textbf{QW}) arithmetic (Hida et al. 2007~\cite{930115})
    \item Quasi Double-Word (\textbf{QDW}) arithmetic (Pair Arithmetic (Lange and Rump 2020~\cite{10.1145/3290955}))
    \item Quasi Triple-Word (\textbf{QTW}) arithmetic (Ozaki and Imamura 2023~\cite{weko_231097_1_eng})
    \item Quasi Quadruple-Word (\textbf{QQW}) arithmetic (Ozaki and Imamura 2023~\cite{weko_231097_1_eng})
\end{itemize}
Using FP64 with 53-bit mantissa, DW arithmetic provides approximately 106-bit precision. Similarly, TW arithmetic achieves approximately 159-bit precision (sextuple precision), and QW arithmetic reaches approximately 212-bit precision (octuple precision).
\par

Sparse iterative solvers are a class of computations where high-precision arithmetic can be particularly effective. One motivation is to obtain more accurate solutions; another is to improve convergence. Rounding errors can increase the number of iterations required for convergence, whereas higher precision can mitigate these errors and potentially reduce the iteration count. This may serve as an alternative to preprocessing techniques that are unsuitable for parallel execution. High-precision arithmetic also offers the possibility of faster overall computation. Let $t_{\texttt{LP}}$ and $n_{\texttt{LP}}$ denote the execution time per iteration and the number of iterations to convergence for a low-precision implementation, and $t_{\texttt{MP}}$ and $n_{\texttt{MP}}$ the corresponding values for a higher-precision implementation. Total solution time is reduced when $t_{\texttt{LP}} \times n_{\texttt{LP}} > t_{\texttt{MP}} \times n_{\texttt{MP}}$. Since sparse solvers are typically memory-bound, the additional cost of multi-word arithmetic may be hidden by memory latency, resulting in modest overhead relative to FP64 arithmetic. Consequently, it may be possible to obtain more accurate solutions without a substantial increase in runtime, or even achieve faster execution than FP64.
\par

This study investigates the use of QDW and QTW arithmetic, alongside DW and TW, in sparse iterative solvers. We focus on the Conjugate Gradient (CG) method (Algorithm~\ref{alg:cg}), one of the simplest iterative approaches for solving linear systems ($\bm{Ax} = \bm{b}$) with symmetric positive definite matrices. In quasi algorithms, the trade-off between accuracy and execution time is known to depend on the computational task, yet has not been thoroughly examined. In iterative methods, such accuracy degradation may be amplified by error accumulation across iterations. We develop CG solvers for FP64 problems using QDW and QTW arithmetic, and compare them with FP64, DW, and TW implementations. The evaluation considers three aspects: convergence behavior (iteration count), solution accuracy, and execution time—both per iteration and total time to convergence on x86 CPUs. Based on these results, we discuss the effectiveness of quasi algorithms in terms of the performance–accuracy trade-off.
\par

The remainder of this paper is organized as follows.
Section \ref{sec:related_work} introduces related work.
Section \ref{sec:algorithm} presents the algorithm of QDW and QTW arithmetic.
Section \ref{sec:cg} shows the implementation of multi-word arithmetic and CG solvers.
Section \ref{sec:experiment} presents the experimental results.
Finally, the conclusion is presented in Section \ref{sec:conclusion}. 
\par 

\begin{algorithm}[t]
\caption{\small{CG method for solving $\bm{Ax}=\bm{b}$}. $\bm{x}_0$ is the initial vector. }
\label{alg:cg}
\begin{algorithmic}[1]
\STATE $\bm{p}_0 = \bm{r}_0 = \bm{b} - \bm{Ax}_0$ \hfill // SpMV
\STATE $\rho_{0} = {\bm{r}_{0}}^T\bm{r}_{0}$ \hfill // DOT
\STATE $i=1$
\WHILE{(1)}
	\STATE $\bm{q}_i = \bm{Ap}_i$ \hfill // SpMV
	\STATE $\alpha_i =  \rho_{i} / {\bm{p}_i}^T\bm{q}_i$ \hfill // DOT
	\STATE $\bm{x}_{i+1} = \bm{x}_{i} + \alpha_i \bm{p}_i$ \hfill // AXPY
	\STATE $\bm{r}_{i+1} = \bm{r}_{i} - \alpha_i \bm{q}_i$ \hfill // AXPY
	\STATE $\rho_{i+1} = {\bm{r}_{i+1}}^T\bm{r}_{i+1}$ \hfill // DOT
    \IF{$||\bm{r}_{i+1}||_{2}/||\bm{b}||_{2}<\epsilon$}
    \STATE \textbf{break} 
    \ENDIF
	\STATE $\beta_{i} = \rho_{i+1} / \rho_i$
	\STATE $\bm{p}_{i+1} = \bm{r}_{i+1} + \beta_i \bm{p}_i$ \hfill // SCAL, AXPY
	\STATE $i=i+1$
\ENDWHILE
\end{algorithmic}
\end{algorithm}

\section{Related Work}
\label{sec:related_work}
As noted in Section~\ref{sec:introduction}, various multi-word arithmetic algorithms have been developed to extend mantissa precision, based on Dekker’s Double-Word (DW) arithmetic~\cite{dekker1971}. Examples include Triple-Word (TW)~\cite{10.1109/TC.2019.2918451}, Quadruple-Word (QW)~\cite{930115}, and their quasi variants (QDW~\cite{10.1145/3290955}, QTW, and QQW~\cite{weko_231097_1_eng}), which reduce computational cost at the expense of potential accuracy loss. Multi-word arithmetic is often implemented using FP64 arithmetic to achieve precision beyond FP64. Well-known libraries include QD\footnote{https://github.com/BL-highprecision/QD}, which provides double- and quadruple-word arithmetic in Fortran and C++, and mx\_real\footnote{https://github.com/RIKEN-RCCS/mX\_real}, a C++ implementation supporting both standard and quasi multi-word algorithms. In addition, high- and arbitrary-precision arithmetic libraries based on integer arithmetic have also been developed, including the GNU Multiple Precision Arithmetic Library (GMP)\footnote{https://gmplib.org}~\cite{10.5555/2911024} and the GNU MPFR Library\footnote{https://www.mpfr.org}~\cite{mpfr}. Additionally, both GNU and Intel compilers provide support for FP128 (IEEE binary128), commonly referred to as quadruple precision.
\par 

For linear algebra operations, XBLAS~\cite{xblas} provides BLAS routines employing DW arithmetic on FP64 data. Several studies have also implemented representative BLAS routines on GPUs and evaluated their performance~\cite{MUKUNOKI2020112701,6270805}. MPLAPACK~\cite{mplapack} offers high-precision BLAS and LAPACK based on libraries such as QD and MPFR. Conversely, efforts exist to realize FP64-equivalent computation on systems lacking native FP64 support by implementing multi-word arithmetic using FP32 or lower-precision formats. For instance, double-word arithmetic has been implemented on GPUs without FP64 units~\cite{10.1145/1179622.1179682}, and double- or triple-word arithmetic using bfloat16 (BF16) has also been investigated~\cite{8877427}. The Ozaki scheme~\cite{Ozaki:2012:ETM:2086820.2086827} has been proposed as an alternative approach performing accurate computation at the matrix-multiplication level rather than through high-precision arithmetic. It enables FP64 GEMM using low-precision floating-point~\cite{isc2020mukunoki}\cite{mukunoki2025dgemmfp64arithmetic} or even integer arithmetic~\cite{ootomo2024dgemm,ozaki2025ozakischemeiigemmoriented}.
\par 

For sparse iterative solvers, several studies have explored the use of high-precision arithmetic to improve convergence. In particular, DW-based implementations of quadruple precision have been extensively investigated~\cite{1570854175390407552,Masui2020,AmaneTakei2021,10.1007/978-3-642-55224-3_59}, with some reports indicating reduced solution time due to fewer iterations. Mixed-precision approaches, in which high precision is applied selectively to critical operations, have also been proposed~\cite{399249e463284a95b80257a99a0e19cd}. Furthermore, CG solvers incorporating accurate sparse matrix–vector multiplication (SpMV) and dot products using the Ozaki scheme have been developed~\cite{hpcasia2021mukunoki}.
\par 

With respect to quasi algorithms, accuracy assessments have been conducted for matrix multiplication and Cholesky decomposition~\cite{weko_231097_1_eng}. However, their applicability to iterative solvers has not been evaluated. Because iterative computations are prone to error accumulation, the accuracy loss inherent in quasi algorithms may have a greater impact in this context. Moreover, their performance has not yet been investigated.
\par 

\section{Multi-Word Arithmetic}
\label{sec:algorithm}
This section presents the algorithms for quasi multi-word arithmetic. The CG method requires addition, multiplication, and division. Among these, we describe the addition and multiplication algorithms used by SpMV and BLAS routines, which account for the majority of execution time.
\par 

Hereafter, $\texttt{fl}(\cdots)$ denotes that all operations within the parentheses are performed using FP64 arithmetic with round-to-nearest-even rounding. $\texttt{fma}(\cdots)$ denotes computation using the FP64 fused multiply–add (FMA) operation ($a\times b+c$). $u$ represents the unit round-off for FP64 ($u = 2^{-53}$). The algorithm's operation count is based on FP64 floating-point operations, with FMA counted as one operation. It is assumed that in FMA, $a\times b-c$ is also performed as one operation without requiring a sign-reversal instruction.
\par 

\begin{algorithm}[H]
    \caption{$[x,y] = $\texttt{TwoSum} $(a, b)$}
    \label{alg:twosum}
    \begin{algorithmic}[1]
    \STATE $x \leftarrow$ \texttt{fl}($a + b$)
    \STATE $z \leftarrow$ \texttt{fl}($x - a$)
    \STATE $y \leftarrow$ \texttt{fl}($(a-(x-z))+(b-z)$)
\end{algorithmic}
\end{algorithm}
\vspace{-0.5em}
\begin{algorithm}[H]
    \caption{$[x,y] = $\texttt{QuickTwoSum} $(a, b)$}
    \label{alg:quicktwosum}
    \begin{algorithmic}[1]
    \STATE $x \leftarrow$ \texttt{fl}($a + b$)
    \STATE $y \leftarrow$ \texttt{fl}($(a-x)+b$)
\end{algorithmic}
\end{algorithm}
\vspace{-0.5em}
\begin{algorithm}[H]
    \caption{$[x,y] = $\texttt{TwoProdFMA} $(a, b)$}
    \label{alg:twoprodfma}
    \begin{algorithmic}[1]
    \STATE $x \leftarrow$ \texttt{fl}($a \times b$)
    \STATE $y \leftarrow$ \texttt{fma}($a \times b -x$)
    \end{algorithmic}
\end{algorithm}
\vspace{-0.5em}
\begin{algorithm}[H]
    \caption{$[c1, c2] = $\texttt{DWadd} $(a1, a2, b1, b2)$}
    \label{alg:dwadd}
    \begin{algorithmic}[1]
    \STATE $[s, e] \leftarrow \texttt{TwoSum}(a1, b1)$
    \STATE $e \leftarrow$ \texttt{fl}($e + a2 + b2$)
    \STATE $[c1, c2] \leftarrow$ \texttt{QuickTwoSum}($s, e$)
\end{algorithmic}
\end{algorithm}
\vspace{-0.5em}
\begin{algorithm}[H]
    \caption{$[c1, c2] = $\texttt{DWmul} $(a1, a2, b1, b2)$}
    \label{alg:dwmul}
    \begin{algorithmic}[1]
    \STATE $[p, e] \leftarrow $ \texttt{TwoProdFMA}($a1, b1$)
    \STATE $e \leftarrow $ \texttt{fma}($a1 \times b2 + e$)
    \STATE $e \leftarrow $ \texttt{fma}($a2 \times b1 + e$)
    \STATE $[c1, c2] \leftarrow $ \texttt{QuickTwoSum}($p, e$)
\end{algorithmic}
\end{algorithm}
\vspace{-0.5em}
\begin{algorithm}[H]
    \caption{$[c1, c2, c3] = $\texttt{QTWadd} $(a1, a2, a3, b1, b2, b3)$}
    \label{alg:qtwadd}
    \begin{algorithmic}[1]
    \STATE $[c1, e1] \leftarrow$ \texttt{TwoSum}($a1, b1$)
    \STATE $[c2, e2] \leftarrow$ \texttt{TwoSum}($a2, b2$)
    \STATE $[c2, e3] \leftarrow$ \texttt{TwoSum}($c2, e1$)
    \STATE $c3 \leftarrow$ \texttt{fl}($a3 + b3 + e2 + e3$)
\end{algorithmic}
\end{algorithm}
\vspace{-0.5em}
\begin{algorithm}[H]
    \caption{$[c1, c2, c3] = $\texttt{QTWmul} $(a1, a2, a3, b1, b2, b3)$}
    \label{alg:qtwmul}
    \begin{algorithmic}[1]
    \STATE $[c1, e1] \leftarrow$ \texttt{TwoProdFMA}($a1, b1$)
    \STATE $[t2, e2] \leftarrow$ \texttt{TwoProdFMA}($a1, b2$)
    \STATE $[t3, e3] \leftarrow$ \texttt{TwoProdFMA}($a2, b1$)
    \STATE $[c2, e4] \leftarrow$ \texttt{TwoSum}($t2, t3$)
    \STATE $[c2, e5] \leftarrow$ \texttt{TwoSum}($c2, e1$)
    \STATE $c3 \leftarrow$ \texttt{fl}(\texttt{fma}$(a3 \times b1 + e2)+$\texttt{fma}$(a2 \times b2 + e3)$\\
    \hfill +\texttt{fma}($a1 \times b3 + e4)+e5$)
\end{algorithmic}
\end{algorithm}
\vspace{-0.5em}
\begin{algorithm}[H]
    \caption{$[c1, c2, c3] = $\texttt{VecSum3} $(c1, c2, c3)$}
    \label{alg:vecsum}
    \begin{algorithmic}[1]
    \STATE $[c1, c2] \leftarrow$ \texttt{TwoSum}($c1, c2$)
    \STATE $[c2, c3] \leftarrow$ \texttt{TwoSum}($c2, c3$)
\end{algorithmic}
\end{algorithm}
\vspace{-0.5em}
\begin{algorithm}[H]
    \caption{$[c1, c2, c3] = $\texttt{DxQTWmul} $(a, b1, b2, b3)$}
    \label{alg:qtwmul2}
    \begin{algorithmic}[1]
    \STATE $[c1, e1] \leftarrow$ \texttt{TwoProdFMA}($a, b1$)
    \STATE $[c2, e2] \leftarrow$ \texttt{TwoProdFMA}($a, b2$)
    \STATE $[c2, e5] \leftarrow$ \texttt{TwoSum}($c2, e1$)
    \STATE $c3 \leftarrow$ \texttt{fl}(\texttt{fma}($a \times b3 + e2)+e5$) 
\end{algorithmic}
\end{algorithm}

First, we introduce the error-free transformation algorithms, which form the foundation of multi-word arithmetic. The TwoSum algorithm (Algorithm~\ref{alg:twosum}~\cite{knuth1969}) decomposes $a + b$ into the floating-point result $x = \texttt{fl}(a + b)$ and the corresponding rounding error $y$. The QuickTwoSum algorithm (Algorithm~\ref{alg:quicktwosum}~\cite{knuth1969}) provides an efficient variant, but is valid only when $|a| \geq |b|$. Similarly, TwoProdFMA (Algorithm~\ref{alg:twoprodfma}~\cite{karp1997}) decomposes $a \times b$ into the floating-point result $x = \texttt{fl}(a \times b)$ and its rounding error $y$ using the FMA operation.
\par

For DW arithmetic~\cite{dekker1971}, given $a = a1 + a2$ ($\texttt{fl}(a1+a2)=a1$), $b = b1 + b2$ ($\texttt{fl}(b1+b2)=b1$), and $c = c1 + c2$ ($\texttt{fl}(c1+c2)=c1$), Algorithm~\ref{alg:dwadd} (DWadd) computes an approximation of $a + b$ as $c$, and Algorithm~\ref{alg:dwmul} (DWmul) computes an approximation of $a \times b$ as $c$. The final QuickTwoSum in these algorithms performs normalization, ensuring that the bit ranges of $c1$ and $c2$ do not overlap and that $\texttt{fl}(c1+c2)=c1$ holds. DWadd and DWmul require 11 and 7 operations, respectively.
\par

QDW arithmetic~\cite{10.1145/3290955} simplifies DW arithmetic by omitting this normalization step via QuickTwoSum, thereby allowing overlap between the high and low words. QDWadd and QDWmul require 8 and 4 operations, respectively.
\par

For TW arithmetic, we employ the addition and multiplication algorithms described in the original paper~\cite{10.1109/TC.2019.2918451}. Two multiplication variants -- accurate and fast -- are proposed therein, and we adopt the fast version in this study. Due to space limitations, the algorithm is not reproduced here; however, it follows the same formulation as presented in the original work. TWadd and TWmul require 42+$\alpha$ and 38+$\alpha$ operations, respectively. The counts are taken from the original paper~\cite{10.1109/TC.2019.2918451}. The notation ``+$\alpha$'' corresponds to the cost described as ``test'' in that work, referring to the overhead introduced by branch operations.
\par

For QTW arithmetic~\cite{weko_231097_1_eng}, given $a = a1 + a2 + a3$, $b = b1 + b2 + b3$, and $c = c1 + c2 + c3$, Algorithm~\ref{alg:qtwadd} (QTWadd) computes an approximation of $a + b$ as $c$, while Algorithm~\ref{alg:qtwmul} (QTWmul) computes an approximation of $a \times b$ as $c$. Similar to QDW arithmetic, these algorithms do not enforce $\texttt{fl}(c1+c2)=c1$ and $\texttt{fl}(c2+c3)=c2$. QTWadd and QTWmul require 21 and 24 operations, respectively.
\par

Omitting normalization may degrade accuracy, as repeated operations increase the overlap of bit positions. Therefore, it may be preferable -- or even necessary in the CG method -- to perform normalization periodically. In QDW arithmetic, QuickTwoSum is used for this purpose. In QTW arithmetic, we employ VecSum3 (Algorithm~\ref{alg:vecsum}~\cite{weko_231097_1_eng}), a three-word extension of VecSum~\cite{doi:10.1137/030601818}. Although VecSum3 does not perform strict normalization, it helps mitigate the degree of overlap.
\par

Since our solvers operate on problems defined in FP64, they involve arithmetic between FP64 values and multi-word types. For such cases, we employ algorithms that omit computations on the lower words by assuming those components to be zero. Algorithm~\ref{alg:qtwmul2} provides an example, illustrating multiplication between FP64 and QTW types.
\par

\begin{table*}[t]
\centering
\caption{Matrices $\bm{A}_{\texttt{orig}}$ ($n \times n$). Sorted in descending order of the number of non-zero elements ($n_{nz}$).}
\label{tab:matrix}
\begin{tabular}{r|l|rrrl}
\hline
\#&Matrix ($\bm{A}_{\texttt{orig}}$)&$n$&$n_{nz}$&$n_{nz}/n$&Application\\\hline
1&Hook\_1498&1,498,023&60,917,445&40.67&Structural Problem\\
2&bone010&986,703&47,851,783&48.50&Model Reduction Problem\\
3&nd24k&72,000&28,715,634&398.83&2D/3D Problem\\
4&crankseg\_2&63,838&14,148,858&221.64&Structural Problem\\
5&crankseg\_1&52,804&10,614,210&201.01&Structural Problem\\
6&nd6k&18,000&6,897,316&40.67&2D/3D Problem\\
7&consph&83,334&6,010,480&72.13&2D/3D Problem\\
8&pdb1HYS&36,417&4,344,765&119.31&Weighted Undirected Graph\\
\hline
\end{tabular}
\end{table*}

\section{Implementation of Multi-word Arithmetic and CG Solvers}
\label{sec:cg}

\subsection{Multi-word Type and Arithmetic}
The two-word types used for DW and QDW are stored in a structure consisting of two FP64 values, while the three-word types used for TW and QTW consist of three FP64 values. Arrays of these types are allocated in an Array of Structures (AoS) format. Arithmetic operations are implemented as inline functions and SIMD-vectorized using AVX2 and AVX-512 intrinsics. In the TW algorithm, conditional branching occurs within individual arithmetic operations, which complicates SIMD implementation; therefore, SIMD vectorization is not applied to branches within TW. By contrast, the QTW algorithm eliminates such conditional dependencies, making it more amenable to vectorization.
\par 

\subsection{SpMV and Vector Operations}
Sparse matrix–vector multiplication (SpMV) and vector operations (DOT, AXPY, and SCAL) are parallelized using both OpenMP and SIMD (in FP64 implementation). For OpenMP parallelization in SpMV, we adopt a simple approach based on one-dimensional block partitioning of the output vector. The Compressed Sparse Row (CSR) format is used for sparse matrix storage. Although the CG method involves symmetric matrices, no symmetry-specific optimizations are applied; the matrices are converted to general form prior to computation.
\par

\subsection{CG Solvers}
We implement CG solvers using DW, QDW, TW, and QTW arithmetic. For comparison, a baseline FP64 implementation is also provided. In all implementations, the coefficient matrix $\bm{A}$ and right-hand side vector $\bm{b}$ are given in FP64, while the solution vector $\bm{x}$ is computed in the format corresponding to the selected arithmetic. All vectors and scalar variables within the CG method are likewise stored in their respective arithmetic types. However, the computation of the relative residual norm (line 10 in Algorithm~\ref{alg:cg}), which does not influence convergence, is performed in FP64 across all implementations.\par 

\subsection{Normalization}
As discussed in the previous section, quasi algorithms do not apply normalization after each arithmetic operation, which can lead to accuracy degradation if computations proceed unchecked. In our preliminary experiments (Section~\ref{sec:normalization}), no convergence was achieved without normalization. However, when normalization was applied to the residual vector $\bm{r}$ after the AXPY operation in line 8 of Algorithm~\ref{alg:cg}, convergence was obtained in approximately the same, or even fewer, iterations as non-quasi algorithms. Since $\bm{r}$ is updated iteratively in CG, this point is a natural location for normalization. Applying it once per iteration, immediately after AXPY, incurs minimal overhead, as SpMV typically dominates the computational cost in CG. Unless otherwise noted, the QDW and QTW implementations normalize at this location. We note, however, that applying normalization more frequently may improve accuracy, albeit at the cost of increased execution time; thus, the optimal frequency and placement of normalization remains an open question.
\par 

\begin{figure*}[t]
\centering
\includegraphics[width=\hsize]{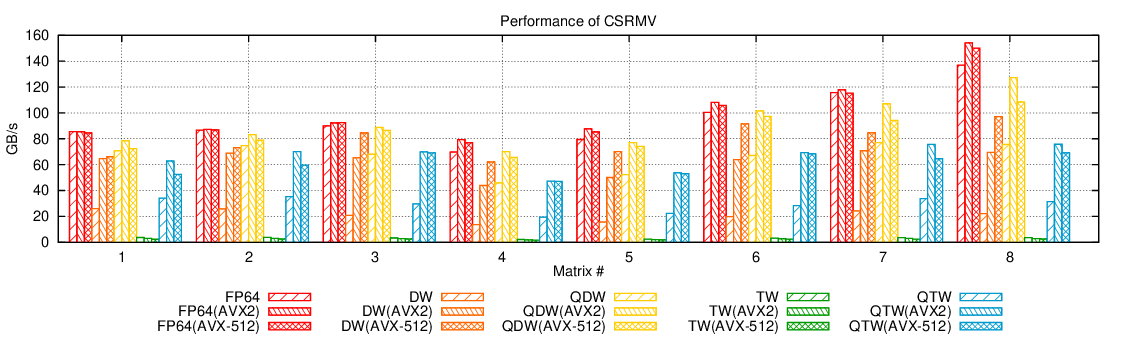}
\caption{Performance of SpMV in GB/s. Note: The matrix is stored in FP64 in all cases, while the vectors are stored in the format corresponding to the arithmetic used.} 
\label{fig:spmv}
\end{figure*}

\begin{figure}[t]
\centering
\includegraphics[width=\hsize]{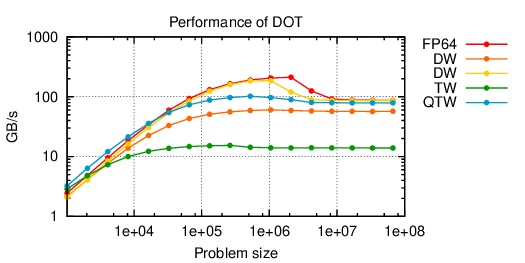}
\caption{Performance of DOT in GB/s.} 
\label{fig:dot}
\end{figure}

\begin{figure}[t]
\centering
\includegraphics[width=\hsize]{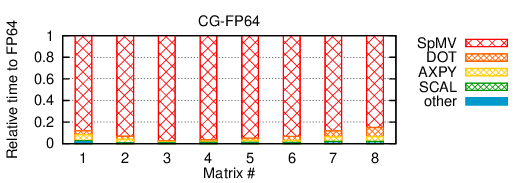}
\includegraphics[width=\hsize]{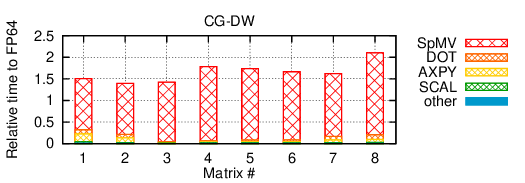}
\includegraphics[width=\hsize]{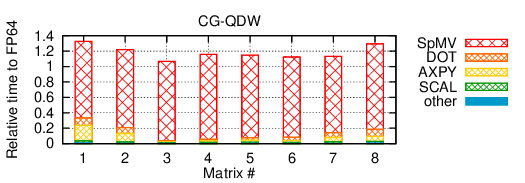}
\includegraphics[width=\hsize]{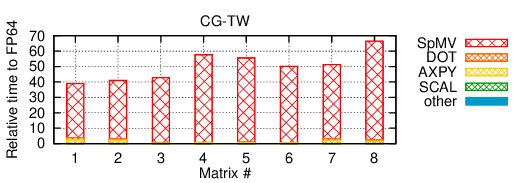}
\includegraphics[width=\hsize]{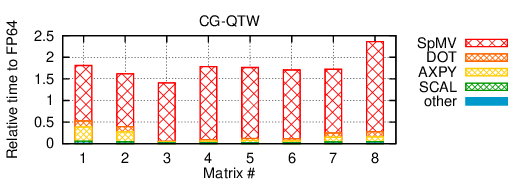}
\caption{Breakdown of relative execution time to FP64 for 100 iterations.} 
\label{fig:cg}
\end{figure}

\section{Numerical Experiments}
\label{sec:experiment}

\subsection{Experimental Conditions}
As the evaluation environment, we use a system\footnote{One node of the supercomputer ``Flow'' cloud system at Nagoya University.} equipped with an Intel Xeon Gold 6230 CPU (Cascade Lake, 20 cores, 2.10–3.90 GHz, 1344 GFlop/s in FP64) and 16 GB of DDR4 memory (2933 MHz, 140.784 GB/s with six memory channels per socket). The machine adopts a Non-Uniform Memory Access (NUMA) architecture with four CPUs, and we configure numactl to use only one socket\footnote{\texttt{numactl -physcpubind=0-19 -membind=0}}. The operating system is CentOS Linux release 7.7.1908 (kernel 3.10.0-1062.9.1.el7.x86\_64), and the code is compiled using \texttt{g++ 11.3.0} with the options \texttt{-O3 -march=native -fopenmp}. One thread is assigned per core.
\par 

For problem generation, we use a method that produces systems with known exact solutions~\cite{published_papers/15365717}. Given an original matrix $\bm{A}{\texttt{orig}}$ and a true solution $\bm{x}^{}$, this method constructs a perturbed matrix $\bm{A}$ and a right-hand side vector $\bm{b}$ such that $\bm{A}\bm{x}^{} = \bm{b}$ holds exactly. The true solution is set to $\bm{x}^{*} = [1, 1, \ldots, 1]^{T}$. Using this approach, solution accuracy is evaluated using the relative error norm $||\bm{x}_{k} - \bm{x}^{}||_{2} / ||\bm{x}^{}||_{2}$. Although the true relative residual norm $||\bm{b} - \bm{A}\bm{x}_{k}||_{2} / ||\bm{b}||_{2}$ is commonly used, the relative error norm provides a stricter assessment of accuracy. The initial vector for the CG method is set to $\bm{x}_{0} = \bm{0}$. To assess the capability of high-precision arithmetic, we evaluate convergence under three tolerances: $\epsilon = 10^{-16}$, $10^{-24}$, and $10^{-32}$.
In this study, we use eight symmetric positive definite matrices (Table~\ref{tab:matrix}) from the SuiteSparse Matrix Collection~\cite{10.1145/2049662.2049663} as $\bm{A}_{\texttt{orig}}$. These matrices are intentionally selected to highlight the reduction in iteration count achieved with high-precision arithmetic.
\par 

\begin{figure*}[t]
\begin{minipage}[b]{0.49\hsize}
\centering
\includegraphics[width=\hsize]{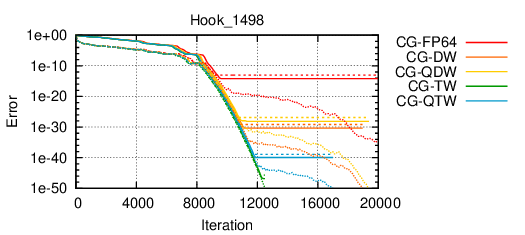}
\includegraphics[width=\hsize]{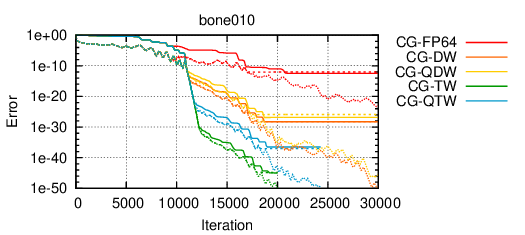}
\includegraphics[width=\hsize]{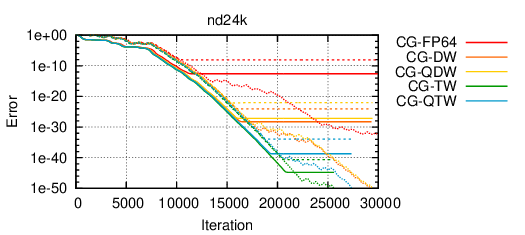}
\includegraphics[width=\hsize]{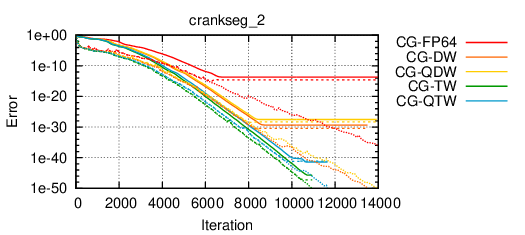}
\end{minipage}
\hfill
\begin{minipage}[b]{0.49\hsize}
\centering
\includegraphics[width=\hsize]{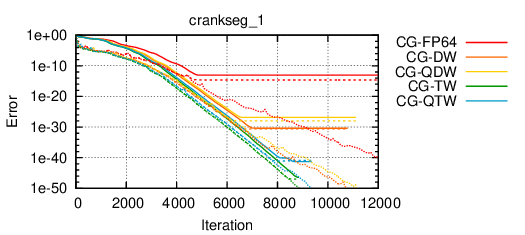}
\includegraphics[width=\hsize]{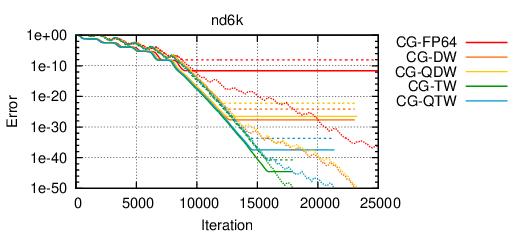}
\includegraphics[width=\hsize]{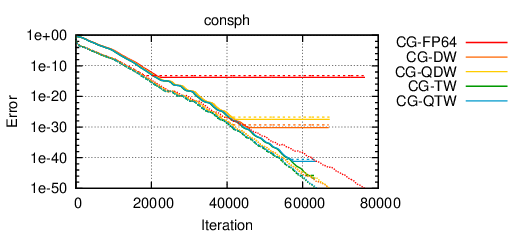}
\includegraphics[width=\hsize]{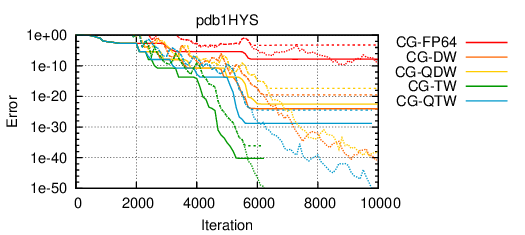}
\end{minipage}
\caption{
Convergence history, plotted every 100 iterations.
Solid lines: relative error norm ($||\bm{x}_{k}-\bm{x}^{*}||_{2}/||\bm{x}^{*}||_{2}$);
dash-dotted lines: true relative residual norm ($||\bm{b}-\bm{Ax}_k||_{2}/||\bm{b}||_{2}$); dotted lines: relative residual norm ($||\bm{r}_k||_{2}/||\bm{b}||_{2}$).} 
\label{fig:cg_conv}

\end{figure*}

\subsection{Results}
\label{sec:preex}

\subsubsection{Throughput of SpMV and DOT}
Fig.~\ref{fig:spmv} shows the throughput of SpMV in GB/s (best result among 10 runs) without SIMD, with AVX2, and with AVX-512. Recall that the matrix is stored in FP64, while the vectors use the format corresponding to the arithmetic. Thus, if performance is memory-bound, similar throughput is expected regardless of the arithmetic type. However, algorithms with higher computational cost become compute-bound and consequently exhibit lower throughput. Matrices with smaller dimensions are more likely to benefit from cache hits, resulting in higher performance. The impact of SIMD acceleration is limited for FP64 and QDW, as their performance is already memory-bound even without vectorization. In contrast, DW, QTW, and TW are compute-bound in non-SIMD form but can become memory-bound after SIMD vectorization. Moreover, TW includes branches that hinder vectorization, leading to significantly reduced performance. The performance difference between AVX2 and AVX-512 varies across matrices; however, AVX2 generally achieved the best performance and is therefore used in subsequent evaluations (for SpMV and other vector operations).
\par 

Fig.~\ref{fig:dot} shows the throughput of the DOT operation with AVX2 vectorization (best of 100 runs). Except for TW, the performance converges to approximately 90~GB/s for sufficiently large problem sizes that no longer fit in cache.
\par 

\subsubsection{Performance of CG Solvers}
The implementations used in the experiments are denoted as follows: CG-FP64, CG-DW, CG-QDW, CG-TW, and CG-QTW, corresponding to implementations using FP64, DW, QDW, TW, and QTW arithmetic, respectively. 
\par 

Fig.~\ref{fig:cg} shows the relative execution time -- normalized to FP64 -- along with its breakdown over 100 iterations of the CG method. In all cases, SpMV accounts for the majority of the total execution time. Compared to FP64, TW incurs a substantial overhead of up to approximately 67 times, whereas DW incurs about 2.1 times, QDW about 1.3 times, and QTW about 2.4 times overhead, with much of their computational cost masked by memory access latency.
\par 

Fig.~\ref{fig:cg_conv} shows the convergence history for iterations up to $\epsilon = 10^{-50}$. Three metrics are reported:
\begin{itemize}
\item Relative error norm: $||\bm{x}-\bm{x}^{*}||_{2}/||\bm{x}^{*}||_{2}$ (solid line)
\item True relative residual norm: $||\bm{b}-\bm{Ax}||_{2}/||\bm{b}||_{2}$ (dash-dotted line)
\item Relative residual norm: $||\bm{r}||_{2}/||\bm{b}||_{2}$ (dotted line)
\end{itemize}
All norm calculations above are performed using TW arithmetic. The results show that although the relative residual norm continues to decrease, both the true relative residual norm and the relative error norm stagnate at a certain level. Non-quasi algorithms reach higher accuracy than quasi-algorithms, often at a faster pace.
\par 

Table~\ref{tab:cg} presents the execution time to convergence\footnote{Measured separately from Fig.~\ref{fig:cg_conv}; true relative residual and relative error norms are not computed during these iterations.}, the number of iterations, and the final relative error norm for convergence criteria of $\epsilon = 10^{-16}, 10^{-24},$ and $10^{-32}$. For each problem, the best result among all implementations is underlined, although in some cases the margin over the next best result is negligible. When high-precision arithmetic reduces the number of iterations, the execution-time overhead relative to FP64 becomes smaller than that shown in Fig.~\ref{fig:cg}. Furthermore, QTW arithmetic substantially reduces execution time compared to TW, while achieving accuracy comparable to TW at a cost not much higher than DW or QDW. This clearly demonstrates the effectiveness of QTW.
\par 

\begin{table*}[t]
\centering
\caption{
Execution time to convergence, number of iterations (\#iter), and relative error norm (err, $||\bm{x}-\bm{x}^{*}||_{2}/||\bm{x}^{*}||_{2}$) are reported. The best results among CG-FP64, CG-DW, CG-QDW, CG-TW, and CG-QTW are underlined.
}
\label{tab:cg}
(a) $\epsilon=10^{-16}$\\
\setlength{\tabcolsep}{5pt}
\begin{tabular}{r|rrr|rrr|rrr|rrr|rrr}
\hline
\#&
\multicolumn{3}{|c|}{CG-FP64}&\multicolumn{3}{|c|}{CG-DW}&\multicolumn{3}{|c}{CG-QDW}&\multicolumn{3}{|c}{CG-TW}&\multicolumn{3}{|c}{CG-QTW}\\
&sec&\#iter&err&sec&\#iter&err&sec&\#iter&err&sec&\#iter&err&sec&\#iter&err\\\hline
1&\underline{1.0e+02}&9638&5.8e-15&1.4e+02&9328&\underline{1.0e-15}&1.3e+02&9375&1.1e-15&3.7e+03&\underline{9122}&1.1e-15&1.8e+02&9204&1.1e-15\\
2&1.7e+02&21780&3.6e-13&1.3e+02&11645&2.7e-15&\underline{1.2e+02}&12547&6.9e-15&3.5e+03&\underline{11349}&7.5e-17&1.5e+02&11352&\underline{7.4e-17}\\
3&6.0e+01&14708&4.8e-13&7.3e+01&13236&\underline{1.5e-19}&\underline{5.8e+01}&13246&1.6e-19&2.2e+03&\underline{12972}&1.6e-19&7.6e+01&13057&1.7e-19\\
4&1.5e+01&6504&8.2e-14&2.1e+01&5193&\underline{6.9e-14}&\underline{1.4e+01}&5274&7.4e-14&6.0e+02&\underline{4675}&7.8e-14&2.0e+01&4837&7.2e-14\\
5&\underline{7.6e+00}&4782&1.3e-13&1.1e+01&4076&7.3e-14&7.6e+00&4140&\underline{7.1e-14}&3.2e+02&\underline{3796}&7.6e-14&1.1e+01&3891&7.3e-14\\
6&\underline{9.4e+00}&11233&2.2e-12&1.5e+01&10557&\underline{9.8e-20}&9.9e+00&10574&1.0e-19&4.3e+02&\underline{10334}&1.1e-19&1.5e+01&10409&1.0e-19\\
7&\underline{1.7e+01}&21671&\underline{3.7e-14}&2.5e+01&21100&3.7e-14&1.9e+01&21137&3.7e-14&8.0e+02&\underline{20646}&3.9e-14&2.8e+01&20735&3.9e-14\\
8&5.4e+00&12807&1.1e-08&5.0e+00&5777&1.0e-24&\underline{3.4e+00}&6285&2.9e-23&1.2e+02&\underline{4403}&9.2e-23&4.9e+00&5346&\underline{4.5e-26}\\
\hline
\end{tabular}
\vspace{0.2em}
\\(b) $\epsilon=10^{-24}$\\
\setlength{\tabcolsep}{5pt}
\begin{tabular}{r|rrr|rrr|rrr|rrr|rrr}
\hline
\#&
\multicolumn{3}{|c|}{CG-FP64}&\multicolumn{3}{|c|}{CG-DW}&\multicolumn{3}{|c}{CG-QDW}&\multicolumn{3}{|c}{CG-TW}&\multicolumn{3}{|c}{CG-QTW}\\
&sec&\#iter&err&sec&\#iter&err&sec&\#iter&err&sec&\#iter&err&sec&\#iter&err\\\hline
1&1.6e+02&15992&5.4e-15&1.6e+02&10329&8.6e-24&\underline{1.4e+02}&10389&8.6e-24&4.1e+03&\underline{10109}&\underline{8.0e-24}&2.0e+02&10201&8.2e-24\\
2&2.6e+02&33372&3.2e-13&1.7e+02&15801&7.4e-23&\underline{1.6e+02}&16570&2.3e-21&3.7e+03&\underline{11854}&\underline{7.2e-25}&1.6e+02&11859&3.8e-23\\
3&8.7e+01&21870&5.3e-13&8.8e+01&15871&\underline{1.0e-27}&\underline{6.9e+01}&15893&1.3e-27&2.6e+03&\underline{15570}&1.0e-27&9.2e+01&15672&1.1e-27\\
4&2.1e+01&9149&2.5e-14&2.8e+01&6902&6.1e-22&\underline{1.9e+01}&7033&6.6e-22&8.0e+02&\underline{6217}&6.6e-22&2.6e+01&6426&\underline{6.0e-22}\\
5&1.1e+01&7049&8.2e-14&1.5e+01&5434&\underline{6.0e-22}&\underline{1.0e+01}&5526&6.7e-22&4.3e+02&\underline{5044}&6.5e-22&1.4e+01&5179&6.6e-22\\
6&1.5e+01&17934&1.4e-12&1.7e+01&12469&8.7e-28&\underline{1.2e+01}&12493&2.9e-27&5.1e+02&\underline{12217}&\underline{8.2e-28}&1.8e+01&12297&9.3e-28\\
7&\underline{2.7e+01}&35183&1.8e-14&4.1e+01&34286&\underline{3.4e-22}&3.0e+01&34341&3.5e-22&1.3e+03&\underline{33555}&3.6e-22&4.6e+01&33716&3.5e-22\\
8&7.9e+00&18317&7.3e-09&5.8e+00&6712&1.2e-24&\underline{3.7e+00}&6886&3.4e-23&1.3e+02&\underline{4797}&\underline{1.2e-33}&5.2e+00&5712&1.1e-29\\
\hline
\end{tabular}
\vspace{0.2em}
\\(c) $\epsilon=10^{-32}$\\
\setlength{\tabcolsep}{5pt}
\begin{tabular}{r|rrr|rrr|rrr|rrr|rrr}
\hline
\#&
\multicolumn{3}{|c|}{CG-FP64}&\multicolumn{3}{|c|}{CG-DW}&\multicolumn{3}{|c}{CG-QDW}&\multicolumn{3}{|c}{CG-TW}&\multicolumn{3}{|c}{CG-QTW}\\
&sec&\#iter&err&sec&\#iter&err&sec&\#iter&err&sec&\#iter&err&sec&\#iter&err\\\hline
1&1.9e+02&18661&5.4e-15&1.7e+02&11181&4.6e-31&\underline{1.6e+02}&11739&9.6e-29&4.4e+03&\underline{10937}&\underline{5.5e-32}&2.2e+02&11039&5.6e-32\\
2&3.6e+02&46291&3.7e-13&2.0e+02&18939&3.0e-29&\underline{1.9e+02}&19066&4.5e-28&3.8e+03&\underline{12326}&\underline{2.0e-31}&2.1e+02&16012&3.6e-30\\
3&1.1e+02&28112&7.9e-13&1.1e+02&19838&2.7e-29&\underline{9.2e+01}&21279&5.6e-28&3.0e+03&\underline{18070}&\underline{9.5e-36}&1.1e+02&18191&9.8e-36\\
4&2.9e+01&12631&4.8e-14&3.6e+01&8561&8.6e-30&\underline{2.4e+01}&8714&2.0e-28&1.0e+03&\underline{7709}&6.1e-30&3.3e+01&7964&\underline{5.9e-30}\\
5&1.5e+01&9588&7.1e-14&1.8e+01&6771&6.4e-30&\underline{1.3e+01}&6871&1.6e-27&5.4e+02&\underline{6295}&\underline{6.0e-30}&1.7e+01&6453&6.1e-30\\
6&1.8e+01&21987&7.4e-13&2.1e+01&15079&1.8e-28&\underline{1.4e+01}&15121&3.0e-27&5.8e+02&\underline{13950}&9.0e-36&2.0e+01&14056&\underline{8.6e-36}\\
7&\underline{3.7e+01}&48086&1.8e-14&5.4e+01&44710&\underline{2.6e-30}&4.0e+01&44790&3.2e-28&1.7e+03&\underline{43768}&2.7e-30&5.9e+01&43972&2.7e-30\\
8&1.1e+01&25802&1.1e-08&6.4e+00&7326&1.1e-24&\underline{4.1e+00}&7532&3.7e-23&1.4e+02&\underline{5294}&\underline{1.4e-40}&5.7e+00&6331&1.3e-29\\
\hline
\end{tabular}
\end{table*}

\subsubsection{Normalization}
\label{sec:normalization}
In our implementations of the quasi algorithms, normalization is applied when updating the residual vector using AXPY. Omitting this step, however, results in non-convergence. Figure~\ref{fig:normalization} presents the convergence histories for consph and pdb1HYS when no normalization is applied in CG-QDW and CG-QTW, denoted as CG-QDW-NN and CG-QTW-NN, respectively.
\par 

\begin{figure}[t]
\centering
\includegraphics[width=\hsize]{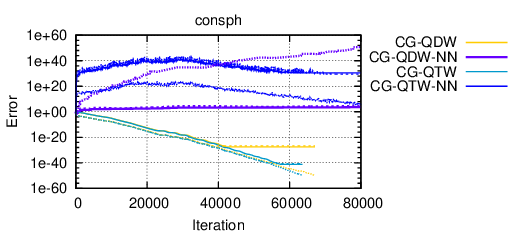}
\includegraphics[width=\hsize]{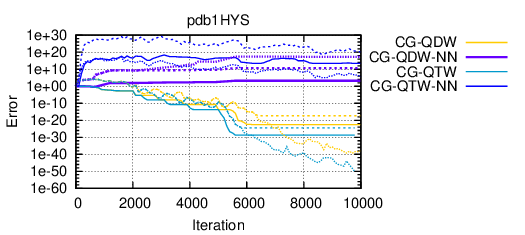}
\caption{Convergence history without normalization. Line styles correspond to those in Fig.~\ref{fig:cg_conv}.} 
\label{fig:normalization}
\end{figure}

\section{Conclusion}
\label{sec:conclusion}
This paper examined double-word (DW) and triple-word (TW) arithmetic, along with their quasi variants QDW and QTW, to improve solution accuracy and convergence in CG solvers. Both runtime performance and numerical accuracy were evaluated. Although quasi algorithms incur accuracy degradation compared to non-quasi variants—primarily due to error accumulation in consecutive operations—this can be mitigated in iterative solvers by applying normalization once per iteration to the residual vector. Nevertheless, quasi algorithms reduce execution time compared to FP64 implementations based on conventional multi-word arithmetic. In particular, QTW significantly lowers execution time relative to TW due to its reduced computational cost and SIMD-friendly structure, providing a lightweight approach to achieving higher-precision solutions than DW and QDW at minimal additional cost.
\par 

Future work includes several directions. First, normalization in quasi algorithms entails a trade-off between accuracy and execution time. While it was applied once per iteration here, more frequent insertion may improve accuracy, warranting further investigation. Second, this study focused on basic unpreconditioned CG; extending the evaluation to other iterative methods, incorporating preconditioning, and exploring mixed-precision strategies are important directions. Third, in distributed environments, communication latency often dominates performance, making the overhead of multi-word arithmetic relatively less significant. This suggests potential speedups through iteration reduction or by omitting preprocessing. Finally, with the growing demand for AI computing, AI-oriented processors offering limited FP64 performance -- or lacking FP64 support entirely -- have emerged. In such environments, implementing multi-word arithmetic using low-precision formats (e.g., FP16 or FP32) may provide a viable means of compensating for FP64 performance limitations.
\par 


\section*{Acknowledgment}
This research was supported by JSPS KAKENHI Grant \#23H03410 and \#25K24387, as well as by the Joint Usage/Research Center for Interdisciplinary Large-scale Information Infrastructures (JHPCN) and the High-Performance Computing Infrastructure (HPCI) under project \#jh250015.

\bibliographystyle{IEEEtran}
\bibliography{poat_qtw_cg}

\end{document}